\documentclass[12pt,a4paper]{article}

\usepackage{natbib}
\bibpunct{(}{)}{;}{a}{}{,}
\usepackage{graphicx}
\usepackage{revsymb}    
\usepackage{amsmath}  
\usepackage{amssymb}  
\usepackage[usenames]{color}    
\usepackage{epstopdf}   
\newcommand{\vt}{{\rm v}}

\newcommand{\derivp} [2] {\frac {\partial #1 } {\partial #2} }
\newcommand{\deriv} [2] {\frac {\textrm{d} #1 } {\textrm{d} #2} }

\newcommand{\eq}[1] {Eq.\,(\ref{#1})}

\usepackage{times}
\title{\bf Transport of angular momentum by waves in stars}

\author{Kevin Belkacem$^1$\\
	\vspace{0.5cm}\\
	\normalsize $^1$  LESIA, Observatoire de Paris, CNRS, Universit\'e PSL, Sorbonne Universit\'e, \\ 
	\normalsize Univ. Paris Diderot, Sorbonne Paris Cit\'e, 5 place Jules Janssen, 92195 Meudon, France}

\date{\mbox{}}
\begin{document}
\maketitle
\setcounter{page}{1001}
\pagestyle{plain}
    \makeatletter
    \renewcommand*{\pagenumbering}[1]{%
       \gdef\thepage{\csname @#1\endcsname\c@page}%
    }
    \makeatother
\pagenumbering{arabic}

%
% WE REDEFINE THE plain LaTeX PAGESTYLE !!! 
% THIS PAGESTYLE WILL BE USED FOR THE FIRST PAGE ONLY !
% Please do not change the following lines
%
\def\bull{\vrule height .9ex width .8ex depth -.1ex}
\makeatletter
\def\ps@plain{\let\@mkboth\gobbletwo
\def\@oddhead{}\def\@oddfoot{\hfil\scriptsize\bull\quad
"How Much do we Trust Stellar Models?", held in Li\`ege (Belgium), 10-12 September 2018 \quad\bull}%
\def\@evenhead{}\let\@evenfoot\@oddfoot}
\makeatother
%
% AND DEFINE OUR MACROS FOR THE REFERENCE LIST
% I.E \beginrefer \refer and \endrefer
%
\def\beginrefer{\section*{References}%
\begin{quotation}\mbox{}\par}
\def\refer#1\par{{\setlength{\parindent}{-\leftmargin}\indent#1\par}}
\def\endrefer{\end{quotation}}
%
% BEGIN THE ABSTRACT WITH \noindent\small, ENCLOSE IT IN A GROUP
% AND BOLDFACE THE TITLE.
%
{\noindent\small{\bf Abstract:} 
Transport of angular momentum is a long-standing problem in stellar physics which recently became more acute thanks to the observations of the space-borne mission \emph{Kepler}. Indeed, the need for an efficient mechanism able to explain the rotation profile of low-mass stars has been emphasized by asteroseimology and waves are among the potential candidates to do so. 
In this article, our objective is not to review all the literature related to the transport of angular momentum by waves but rather to emphasize the way it is to be computed in stellar models. We stress that to model wave transport of angular momentum is a non-trivial issue that requires to properly account for interactions between meridional circulation and waves. Also, while many authors only considered the effect of the wave momentum flux in the mean momentum equation, we show that this is an incomplete picture that prevents from grasping the main physics of the problem. We thus present the Transform Eulerian Formalism (TEM) which enable to properly address the problem. 
}
\vspace{0.5cm}\\
% SPECIFY UP TO 5 KEYWORDS SEPARATED BY ' -- '
\section{Introduction} 

Rotation has  fundamental  effects  on  stellar  evolution. For instance, it  induces meridional circulations, as  well  as  shear and baroclinic instabilities,  which contribute to  the  redistribution of angular momentum and to the mixing of chemical elements (e.g. Maeder 2009). Therefore, it is essential to properly understand and model the physical processes responsible for the transport and redistribution of angular momentum. Among them, waves\footnote{Note that by waves we denote both  progressive waves and normal modes.} have been identified to play a major role. Indeed, the wave/mean-flow interactions has been considered for a long time for geophysical flows and more precisely the interaction between waves, angular momentum, and meridional circulation. This had been extensively studied in the 60s and 70s in the context of middle atmosphere dynamics (see for instance Andrews et al. 1987 and Holton 1992 for extensive reviews). All those efforts have permitted to unveil and clarify the nature of these interactions and led to the developments of adapted formalisms among which the Transformed Eulerian Mean formalism (Andrews \& McIntyre 1976, 1978a) and the Generalized Lagrangian Mean formulation by Andrews \& McIntyre (1978b). For stellar interiors, it has been first addressed from a theoretical point of view with the pioneer works of Press (1981) and Ando (1983). Both authors were mainly interested in addressing the problem of the redistribution of angular momentum by waves (even if with quite different motivations) and thus investigated the question of the interaction between waves and rotation through wave momentum stresses in the mean angular momentum equation. Subsequently, two set of distinct works had been developed almost in parallel. 

On one had, following Ando (1983), efforts have been made to address the problem of the redistribution of angular momentum in massive stars by unstable normal modes (e.g. Ando1986, Lee \& Saio 1993, Lee 2007, Townsend \& MacDonald 2008, Ishimatsu \& Shibahashi 2013, Townsend 2014, Lee et al.  2014, Townsend 2018). As an illustration, one of the main motivations was to explain the episodic mass loss in Be stars. The more studied scenario can be summarized as follows; angular momentum is deposited at the surface by normal modes. Thus, the rotation locally increases to reach the break-up limit and then allows for mass loss. However, this scenario is not commonly accepted and, for instance,  was questioned by Ishimatsu \& Shibahashi (2013) and Shibahashi (2014) because observations show that the break-up velocity is not reached. Consequently, the authors proposed an alternative scenario in which g-modes transfer angular momentum in the stellar surface, rotation increases (but still below the break-up limit), the critical frequency also increases, thus implying mode leakage and transfer of angular momentum to the disk. Obviously, those scenarios crucially depend on the way angular momentum is deposited at the surface by waves. Therefore, it is mandatory to properly  decipher the interaction between waves and rotation.  

On the other hand, and following Press (1981), many authors have considered the interaction between the internal gravity waves and the mean flow in low-mass stars (e.g. Schatzman 1993; Zahn et al. 1997; Kumar et al. 1999; Charbonnel \& Talon 2005; Mathis et al. 2013). The primary motivation was to explain the quasi-uniform rotation profile in the Solar radiative interior. Internal gravity waves have been shown to be able to extract angular momentum on short time scales (compared to the evolution time-scale) and had then been considered as a serious candidate to explain the solar rotation profile as well as the cool side of the Li-dip 
(e.g. Talon \& Charbonnel 2003,2005; Charbonnel \& Talon 2005). The advent of asteroseismology further strengthened the focus on internal gravity waves as it was a potential candidate for  explaining the weakly increasing rotation contrast of subgiant stars and the spin-down of red giants. It has been shown that they are inefficient for evolved red giants to explain the slow-down of red giants (e.g. Fuller et al. 2014; Pinc\c on et al. 2017) but could explain the rotation profile of subgiants (Pinc\c on et al. 2017).  

Those two set of works, while apparently disconnected,  rely essentially on the same physics which require a deep understanding  of the interaction between the waves and the mean flow. This is, however, not a trivial issue as we will discuss in the following. In particular, we will show that considering the divergence of the wave flux in the momentum equation is not enough and that the energy equation (more precisely the effect of the wave heat flux) is essential to account for as both equations are intricately coupled by meridional circulation. This coupling had however been too often overlooked in stellar physics, while it has been demonstrated to play a fundamental role. Our objective is thus to highlight this issue. To that end, based on the classical azimutaly-averaged equations we will first present in Sect.~\ref{sect2} some arguments to show the need for considering the coupling between the momentum and energy equation.  In Sect.~\ref{sect3}, the Transformed Eulerian Mean formalism will be introduced and an application to mixed modes in red giants will be presented in Sect.~\ref{sect4}. Finally, in Sect.~\ref{sect5}, concluding remarks are provided.

\section{Setting the stage}
\label{sect2}

\subsection{Wave effects on the mean field: heuristic arguments}
\label{heuristic}

In stars, progressive or standing waves are generally considered to have small amplitudes, as (but not only) for solar-like oscillations. This is well justified and naturally motivates to adopt the linear approximation. However, given their  small amplitudes, it is worthwhile to wonder if such waves can have significant effects on the mean flow and are able to modify the rotation profile. To have some hints about this issue, it is very enlightening to use some simple heuristic arguments. To do so, we here essentially follow the discussion as presented by Buhler (2009). Let us first consider the conservation of momentum equation in a schematic form as 
\begin{align}
\label{heuristic_moment}
\derivp{\vec \vt}{t} + \mathcal{L}\left(\vec \vt\right) + \mathcal{B}\left(\vec \vt,\vec \vt\right) = \vec 0  \, , 
\end{align}
where $\vec \vt$ is the velocity field, $\mathcal{L}$ a linear operator, and $\mathcal{B}$ a bi-linear operator. The velocity field can then be expanded using the wave amplitude ($a_w \ll 1$) as a small parameter, thus 
\begin{align}
\label{heuristic_vitesse}
\vec \vt = \vec \vt_0 + a_w \, \vec \vt_1 + a_w^2 \, \vec \vt_2 + O(a_w^3) \, , 
\end{align}
where the first r.h.s term ($\vec \vt_0$) corresponds to the mean flow, the second r.h.s. term ($\vec \vt_1$)  stands for the wave velocity, and the third r.h.s term ($\vec \vt_2$) is the back reaction onto the mean-flow by the waves. Using \eq{heuristic_vitesse} into \eq{heuristic_moment} then provides a hierarchical system of equations. The lowest-order is 
\begin{align}
\derivp{\vec \vt_0}{t} + \mathcal{L}\left(\vec \vt_0\right) + \mathcal{B}\left(\vec \vt_0,\vec \vt_0\right) = \vec 0  \, ,  
\end{align}
which corresponds to the equation governing the mean-flow. At the next order one has
\begin{align}
\derivp{\vec \vt_1}{t} + \mathcal{L}\left(\vec \vt_1\right) + \mathcal{B}\left(\vec \vt_0,\vec \vt_1\right) + \mathcal{B}\left(\vec \vt_1,\vec \vt_0\right) = \vec 0  \, , 
\end{align}
which governs the linear waves. Finally, at the second-order, we obtain
\begin{align}
\label{heuristic_vitesse2}
\derivp{\vec \vt_2}{t} + \mathcal{L}\left(\vec \vt_2\right) + \mathcal{B}\left(\vec \vt_0,\vec \vt_2\right) + \mathcal{B}\left(\vec \vt_2,\vec \vt_0\right) =  -\mathcal{B}\left(\vec \vt_1,\vec \vt_1\right) \, , 
\end{align}
which governs the effect of linear waves onto the mean flow. Then, we immediately note that the back-reaction equation is forced by  wave-related terms of the second order $\mathcal{B}\left(\vec \vt_1,\vec \vt_1\right)$ and that no first-order term appears. This is an important point because, as we will see in Sect.~\ref{sect5},   second-order effects such as Stokes corrections cannot be neglected when considering the effect of waves on the mean flow. 

To go further, let us assume that there is no mean field velocity ($\vec \vt_0 = \vec 0$). Then, \eq{heuristic_vitesse2} reduces to
\begin{align}
\derivp{\vec \vt_2}{t} + \mathcal{L}\left(\vec \vt_2\right)  =  R \, , 
\end{align}
with $R\equiv -\mathcal{B}\left(\vec \vt_1,\vec \vt_1\right)$. 
%, so that the formal solution is 
%\begin{align}
%\vec \vt_2  =  \exp\left(-t \mathcal{L}\right) \vec \vt_2 (0) + \int_{0}^{t} \exp\left(-(t-s) \mathcal{L}\right) R(s) \, {\rm d}s \, .
%\end{align}
Now, assuming that $ \vec \vt_2 (0)= 0$ and that $R$ is independent of time (for steady waves and time-independent mean properties) one has 
\begin{align}
\vec \vt_2  \simeq R \, t = \vec \vt_2 \left(a_w^2 t\right) \, . 
\end{align}
The latter solution can then be interpreted as follows; on a large time-scale, $t=O\left(a_w^{-2}\right)$, we have $\vec \vt_2  = O \left(1\right)$. Otherwise stated, waves can significantly modify the mean flow on long time scales. %Therefore, when estimating the evolution of the rotation profiles in stars, it is essential to assess the effect of the waves. 

\subsection{Azimuthal (or zonal) Eulerian averaging of the primitive equations}

In this section, our objective is to derive the equations that describe the effect of waves on the mean flow or more precisely the interaction between the mean flow and the waves. Thus, we start by considering the continuity, momentum, and energy equations in an inertial frame. They can be written such as
\begin{align}
	\label{continuity}
	&\derivp{\rho}{t} + \vec \nabla \cdot \left( \rho \vec \vt \right) = 0 \\
	\label{moment}
	&\derivp{\left( \rho \vec \vt \right)}{t} + \vec \nabla \cdot \left( \rho \vec \vt \vec \vt \right) = - \vec \nabla p  - \rho \vec \nabla \Phi + \vec X \\
	\label{energy}
	&\derivp{\left( \rho s \right)}{t} + \vec \nabla \cdot \left( \rho \vec \vt s \right) = Q  \, , 
\end{align}
where $\rho$ is the density, $\vec \vt$ the velocity field, $p$ the pressure, $s$ the specific entropy, $\Phi$ the gravitational potential, $\vec X$ is a non-conservative mechanical forcing (e.g., turbulent dissipation), and $Q$ represents heating or cooling terms. Obviously, Eqs.~(\ref{continuity}) to (\ref{energy}) need to be complemented by the appropriate equation of state together with boundary conditions. 

To go further, we need to derive the equation representing the conservation of the  specific angular momentum. This equation is obtained by adopting spherical coordinates $(r,\theta,\phi)$ with the associated unit vectors $(\vec e_r,\vec e_\theta,\vec e_\phi)$ and $\theta=0$ corresponding to the direction of the rotation axis. Then, using the azimuthal component of \eq{moment}, multiplied by $\varpi  \equiv r \sin \theta$, leads to 
\begin{align}
	\label{momentum}
	\derivp{\left( \rho h \right)}{t} + \vec \nabla \cdot \left(  \rho h \vec \vt \right) = - \derivp{p}{\phi} - \rho \derivp{\Phi}{\phi} + \varpi \, X_\phi \, , 
\end{align}
where the specific angular momentum is defined by $h \equiv \varpi \, \vt_\phi $ and $\vt_\phi$ is the azimuthal component of the velocity field. 

To gain insight and tractability to the problem, it is useful to shift from a 3D to a 2D problem and to separate the waves and the mean-flow. To do so, we consider azimuthal (or zonal) average of the equations. Therefore, a given field can be decomposed into a mean part and a perturbation. More precisely, for a given field $A$, we have
\begin{equation}
\label{decomposition_scalar}
A = \overline{A} + A^\prime,
\end{equation}
where $\overline{A}$ is the Eulerian-mean azimuthal average defined by
\begin{equation}
\overline{A} = \frac{1}{2 \pi} \int_0^{2\pi} A \; {\rm d}\phi \, .
\end{equation}
This Eulerian average then possesses the usual properties (additivity, linearity, commutativity with partial differentiation, etc\dots) and $A^\prime$ stands for the non-axisymmetric pertubations. 

Applied to the velocity field, this decomposition permits us to identify the main ingredients of the problem. More precisely, one has 
\begin{align}
\vec \vt = \overline{\vec \vt} + {\vec \vt}^\prime \, ,
\end{align}
where $\overline{\vec \vt} = \overline{\vec \vt}_\bot + \varpi \Omega \, \vec e_\phi$ with $\overline{\vec \vt}_\bot$ corresponding to the meridional circulation. Finally, the non-axisymmetric perturbations are associated with non-axisymmetric waves.

Because we are interested in estimating the effect of waves on the mean flow (and primarily on $\overline{h}$), it is necessary to obtain the mean equations. This is done by applying \eq{decomposition_scalar} to Eqs.~(\ref{continuity}), (\ref{energy}), and (\ref{momentum}) and finally to perform an azimuthal averaging. After some manipulations, and retaining 
 perturbations up to the second order, it gives
\begin{align}
	\label{continuity_eulerian_tmp}
	&\derivp{\overline{\rho}}{t}  
	+ \vec \nabla_\bot \cdot \left( \overline{\rho} \, \overline{\vec \vt}_\bot \right)
	= \mathcal{D} \, , \\
	\label{momentum_eulerian_tmp}
	&\overline{\rho} \derivp{\overline{h}}{t}   
	+ \overline{\rho} \, \left( \overline{\vec \vt}_\bot \cdot \vec \nabla_\bot \right) \overline{h} 
	= - \vec \nabla_\bot \cdot \left( \varpi \overline{\rho} \, \overline{ \vt_\phi^\prime \vec \vt^\prime_\bot} \right) 
	+ \varpi  \overline{X}_\phi + \mathcal{H} \, , \\
	\label{energy_eulerian_tmp}
	&\overline{\rho} \derivp{\overline{s}}{t} + 
	\overline{\rho}\left( \overline{\vec \vt}_\bot \cdot \vec \nabla_\bot \right) \overline{s} 
	= - \vec \nabla_\bot \cdot \left(\overline{\rho} \, \overline{s^\prime \vec \vt^\prime_\bot} \right) + \overline{Q} + \mathcal{S} \, , 
\end{align}
with
\begin{align}
	\label{term_h}
	\mathcal{H} &=  - \overline{\rho^\prime \vec \vt^\prime_\bot} \cdot \vec \nabla_\bot  \overline{h} - \varpi \derivp{\overline{\rho^\prime \vt_\phi^\prime}}{t} - \overline{\rho^\prime \derivp{\Phi^\prime}{\phi}} 
	- \vec \nabla_\bot \cdot \left( \varpi \overline{\rho^\prime \vt_\phi^\prime} \, \overline{\vec \vt}_\bot \right), \\ 
	\label{term_s}
	\mathcal{S} &= -  \overline{\rho^\prime \vec \vt^\prime_\bot} \cdot \vec \nabla_\bot \overline{s}
	- \vec \nabla_\bot \cdot \left(\overline{\rho^\prime s^\prime} \, \overline{\vec \vt}_\bot \right) - \derivp{\overline{ \rho^\prime s^\prime}}{t} ,\\
	\label{term_d}
	\mathcal{D} &= - \vec \nabla_\bot \cdot \left(\overline{\rho^\prime \, \vec \vt^\prime_\bot} \right) ,
\end{align}
and $\vec \nabla_\bot$ and $\vec \vt_\bot$ are the gradient and velocity vector in the meridional plane, respectively. They are defined by
\begin{align}
	\vec \nabla_\bot = \vec e_r  \derivp{}{r}  + \vec e_\theta\frac{1}{r} \derivp{}{\theta}   ,
	\quad {\rm and} \quad \vec \vt_\bot = \vt_r \, \vec e_r + \vt_\theta \, \vec e_\theta \, .
\end{align}

Looking at Eqs.~(\ref{continuity_eulerian_tmp}) to (\ref{energy_eulerian_tmp}), one immediately note that wave-related terms are only of the second order. This is consistent with the heuristic arguments presented in Sect.~\ref{heuristic}. Consequently, it also implies that the wave field can modify the mean-flow only on large time-scales (compared to the wave periods). Moreover, anticipating on the following, one can already note that wave-related second-order effects such as the Stoke drifts cannot be neglected. 

Equations~(\ref{continuity_eulerian_tmp}) to (\ref{energy_eulerian_tmp}) also show that many wave-related terms are potentially able to affect the mean flow. Among them, one can distinguish terms involving a time derivative. They are non-negligible only if the wave amplitude varies with time or if one consider the effect of wave packets. In the latter case those terms are nevertheless expected to be small because of cancellation effects. In contrast, if we consider steady waves (i.e. wave with a constant amplitude), those terms vanish. Such a steady state is often assumed when considering progressive as well as stationary waves in stars. We will thus neglect those terms. 
  
Finally, all terms encompassed into $\mathcal{H}, \mathcal{S}, \mathcal{D}$ exhibit density perturbation ($\rho^\prime$). They are generally considered to be small (e.g. Ando 1983, Unno et al. 1989). This is particularly the case for low-frequency waves ($\sigma_R \ll N$, where $\sigma_R$ is the wave frequency and $N$ the buoyancy frequency) where the anelastic approximation applies (e.g. Dintrans \& Rieutord 2001). Therefore, we neglect the terms $\overline{\rho^\prime \, \vec \vt^\prime_\bot}$, $\overline{\rho^\prime s^\prime}$, and $\overline{\rho^\prime \, \vt_\phi^\prime}$ in Eqs.~(\ref{term_h}) to (\ref{term_d}). In addition, we use the Cowling approximation by neglecting the perturbation of the gravitational potential, so that Eqs.~(\ref{continuity_eulerian_tmp}) to (\ref{energy_eulerian_tmp}) become
\begin{align}
	\label{continuity_eulerian}
	& \derivp{\overline{\rho}}{t}  
	+ \vec \nabla_\bot \cdot \left( \overline{\rho} \, \overline{\vec \vt}_\bot \right)
	= 0 \\
	\label{momentum_eulerian}
	&\overline{\rho} \derivp{\overline{h}}{t}   
	+ \overline{\rho} \, \left( \overline{\vec \vt}_\bot \cdot \vec \nabla_\bot \right) \overline{h} 
	 + \vec \nabla_\bot \cdot \left( \varpi \overline{\rho} \, \overline{ \vt_\phi^\prime \vec \vt^\prime_\bot} \right)
	= \varpi \overline{X}_\phi   \\
	\label{energy_eulerian}
	&\overline{\rho} \derivp{\overline{s}}{t} + 
	\overline{\rho}\left( \overline{\vec \vt}_\bot \cdot \vec \nabla_\bot \right) \overline{s} 
	 + \vec \nabla_\bot \cdot \left(\overline{\rho} \, \overline{s^\prime \vec \vt^\prime_\bot} \right) 
	= \overline{Q}  \, . 
\end{align} 
These equations immediately show that wave forcing of the mean flow is not only due to the wave momentum flux in Eq.~(\ref{momentum_eulerian}) (also sometimes named to as the Reynolds stresses in analogy to the turbulent Reynolds stress) but also to the wave heat flux in the mean entropy equation, Eq. \eqref{energy_eulerian}. It is important to already note that the latter cannot be neglected compared to the former, especially for low-frequency waves. This will be discussed in details in the following sections. 

Moreover, we note that our system of equation is not closed because the mean specific entropy $\overline{s}$ and mean specific angular momentum $\overline{h}$ are not independent. They are connected through the baroclinic equation (also known to as the thermal wind balance equation). It is obtained by taking the curl of the hydrostatic equilibrium equation so that 
\begin{align}
	\label{baroclinic_eulerian}
	&\overline{\rho}^2 \; \vec \nabla_\bot \left(\frac{\overline{h}^2}{\varpi^4} \right) \times \vec \nabla_\bot \varpi + \vec \nabla_\bot \overline{\rho}  \times \vec \nabla_\bot \overline{p} = 0 \, , 
\end{align}
together with the equation of state $\overline{s} = \overline{s} \, (\overline{\rho},\overline{p})$. This equation is fundamental because it determines the balance between the centrifugal (first term of the l.h.s. of Eq.~\ref{baroclinic_eulerian}) and baroclinic (second term of the l.h.s. of Eq.~\ref{baroclinic_eulerian}) torques. Indeed, to ensure the torques balance, the thermal wind equation states that $\Omega(r)$ is completely determined by $\rho(r)$ along an isobar. In other words, $\Omega(r)$ and $\rho(r)$ (therefore $s(r)$)  cannot evolve separately because they are connected by the meridional circulation. 

Finally, in principle solving Eq.~(\ref{continuity_eulerian}) to (\ref{baroclinic_eulerian}) (with the appropriate equation of state and boundary conditions), together with the wave equations, permits one to properly model the wave and mean flow interactions. This is however a non-trivial task. First, the time-scales of the problem are not always commensurable in the sense that we need to solve the wave equations on a short time-scale (typically a dynamical time-scale) while we need to solve mean-equations on a longer time-scale (typically the Kelvin-Helmholtz time-scale). Second, and maybe more importantly, this system of equations made very difficult to make clear what physical mechanism is responsible for the mean-flow driving. This is illustrated by the fact that the second and third terms of the l.h.s. of \eq{momentum_eulerian} and \eq{energy_eulerian} often nearly compensate each others so that the resulting driving of the mean flow is a residual. 

\subsection{Coupling between wave fluxes and meridional circulation: an essential ingredient}

Before going further, it is necessary to insist on a crucial point: \emph{the wave heat flux must not be neglected}, as well as meridional circulation. If neglected, most of the physics of the problem is lost and this potentially leads to incorrect estimates of the effect of waves on the mean-flow and subsequently on the evolution of the rotation profile.   

To figure out the problem, let us consider the conservation of angular momentum and neglect the wave heat flux, meridional circulation, and any external mechanical forcing. After meridional averaging (denoted by $\left< . \right>$), it gives
\begin{align}
\label{momentum_eulerian_absurd}
\left<\overline{\rho} \derivp{\overline{h}}{t}    \right>
=-  \frac{1}{r^2} \derivp{}{r} \left< r^2 \varpi \overline{\rho} \, \overline{ \vt_\phi^\prime \vec \vt^\prime_r} \right>   \, . 
\end{align} 
One then immediately concludes that the wave momentum flux modifies the distribution of angular momentum and thus the rotation profile. Let us now consider adiabatic waves.  If the effect of rotation is neglected then the r.h.s of \eq{momentum_eulerian_absurd} vanishes because $\vt_\phi^\prime$ and $\vt^\prime_r$ are in phase quadrature, thus waves do not modify the rotation profile. However, if we include the effect of rotation onto the waves (even in the adiabatic limit), the r.h.s of \eq{momentum_eulerian_absurd} \emph{does not} vanish and one must conclude that adiabatic waves affect the mean flow. This is indeed a weird conclusion because it would mean that adiabatic waves should be able to modify the angular momentum of the mean flow without energy exchanges. 

To solve this apparent physical issue, one must consider both the wave heat flux and the meridional circulation
\begin{align}
\label{momentum_eulerian_simple}
&\left< \overline{\rho} \derivp{\overline{h}}{t}   \right>
+ \left<\overline{\rho} \, \left( \overline{\vec \vt}_\bot \cdot \vec \nabla_\bot \right) \overline{h}  \right>
= -  \frac{1}{r^2} \derivp{}{r} \left< r^2 \varpi \overline{\rho} \, \overline{ \vt_\phi^\prime \vec \vt^\prime_r} \right>  \\
\label{energy_eulerian_simple}
&\left<\overline{\rho} \derivp{\overline{s}}{t}  \right>+ 
\left<\overline{\rho}\left( \overline{\vec \vt}_\bot \cdot \vec \nabla_\bot \right) \overline{s} \right> 
=-  \frac{1}{r^2} \derivp{}{r} \left<r^2 \overline{\rho} \, \overline{s^\prime \vec \vt^\prime_r} \right>  \, . 
\end{align} 
Still in the limit of adiabatic waves but considering the effect of rotation on the waves, both terms in the r.h.s of \eq{momentum_eulerian_simple} and \eq{energy_eulerian_simple} do not vanish. However, waves do not modify the mean flow as shown by the \emph{non-acceleration theorem} (Andrews and Mclntyre 1976). This apparent paradox is solved by realizing that the wave fluxes of heat and momentum produce a meridional circulation that cancels their tendency to affect the mean flow. It is thus important to stress on several points; 
\begin{itemize}
\item  the heat and momentum wave fluxes do not act independently of each other. 
\item  the connexion is performed by the meridional circulation, which must ensures that $\overline{h}$ and $\overline{s}$ satisfy the baroclinic equation (Eq.~\ref{baroclinic_eulerian}).
\item overlooking the effect of the wave heat flux or the coupling between wave fluxes and meridional circulation does not permit to grasp the main picture of the problem and potentially leads to unphysical or incorrect results. 
\end{itemize}

This non-trivial interplay between the mean-flow and the wave forcing terms was properly addressed decades ago in the context of geophysical flows (e.g. Andrews \& McIntyre 1976,1978a,1978b). In contrast, in the stellar context, the picture is made confused because the wave heat flux is too-often overlooked (e.g., Press 1981, Ando 1983, Lee \& Saio 1993, Zahn et al. 1997, Kumar et al. 1999, Pantillon et al 2007, Mathis 2013, Townsend 2014, Townsend et al. 2018) even if not always (e.g. Lee 2013, Belkacem et al. 2015a,b). In the following sections, we therefore introduce two formalisms, which nicely clarify the physical picture, namely: the Transformed Eulerian Mean (TEM) and the Generalized Lagragian Mean (GLM) formalisms.

\section{Transformed Eulerian Mean (TEM) formalism}
\label{sect3}

As discussed in Sect.~\ref{sect2}, the wave fluxes of heat and momentum produce a meridional circulation which, in the limit of steady and adiabatic waves, cancels their tendency to affect the mean flow. Therefore, the principle of the TEM is to eliminate the advective part of the divergence of the wave heat flux in the entropy equation so as to inject it into the a newly defined meridional circulation that will be named to as the residual circulation. This formalism was introduced by Andrews \& McIntyre (1976,1978a). 

\subsection{Incorporating the divergence of the skew flux into meridional circulation}

The first step consists in splitting the divergence of the wave heat flux into an advective and a diffusive part. To do so, let us consider an isentropic surface with a normal vector define as $\vec n = \vec \nabla_\bot \overline{s} / \vert \vec \nabla_\bot \overline{s} \vert$. Then, the wave heat flux $\vec R = \overline{s^\prime \vec \vt^\prime_\bot}$ can be split into a component along an isentropic surface (the
skew flux) and a component perpendicular to it, such as 
\begin{align}
	\vec R = \left( \vec n \times \vec R \right) \times \vec n + \left(\vec n \cdot \vec R \right) \, \vec n \;. 
\end{align} 
The divergence of the skew flux can then be rewritten as
\begin{align}
	\label{split_div_flux}
	\vec \nabla_\bot \cdot \left[ \left( \vec n \times \vec R \right) \times \vec n \right] &= 
	\vec \nabla_\bot \cdot \left( \frac{  \vec \nabla_\bot \overline{s} \times \vec R }{\vert \vec \nabla_\bot \overline{s} \vert^2} \times \vec \nabla_\bot \overline{s} \right) 
	= \left( \vec \nabla_\bot \times \frac{\vec \nabla_\bot \overline{s} \times \vec R}{\vert \vec \nabla_\bot \overline{s} \vert^2} \right) \cdot \vec \nabla_\bot \overline{s} \nonumber \\
	&= \tilde{\vec \vt}\cdot \vec \nabla_\bot \overline{s}\, ,
\end{align}
where we used the relation $\vec \nabla \cdot (\vec a \times \vec \nabla \alpha) = \vec \nabla \alpha \cdot (\vec \nabla \times \vec a)$, valid for any scalar $\alpha$ and vector $\vec a$. 

Equation~(\ref{split_div_flux}) shows that the skew flux behaves like an advection by the velocity $\tilde{\vec \vt}$. The main motivation underlying the TEM is thus to incorporate the advective part of the wave heat flux into the mean meridional velocity field ($\overline{\vec \vt}$). To do so, the \emph{residual meridional circulation} is thus defined by  
\begin{align}
	\label{residual_v}
	\overline{\rho} \, \overline{\vec \vt}_{\bot}^\dag =  \overline{\rho} \, \overline{\vec \vt}_\bot  + \vec \nabla_\bot \times \left(\overline{\rho} \, \overline{\psi} \, \vec e_\phi\right) \, , 
\end{align}
where the stream function $\overline{\psi}$ is deduced from \eq{split_div_flux} as 
\begin{align}
	\label{def_psi_general}
	\overline{\psi} =  \frac{\vec \nabla_\bot \overline{s} \times \vec R}{\vert \vec \nabla_\bot \overline{s} \vert^2} \cdot \vec e_\phi= \frac{1}{\vert \vec \nabla_\bot \overline{s} \vert^2} \left[ \left(\derivp{\overline{s}}{r}\right) \, \overline{s^\prime \vt_\theta^\prime} - \frac{1}{r} \left(\derivp{\overline{s}}{\theta}\right) \, \overline{s^\prime \vt_r^\prime} \right]  \, .
\end{align}

Now, inserting Eqs.~(\ref{residual_v}) and (\ref{def_psi_general}) into \eq{continuity_eulerian} to \eq{energy_eulerian}, we have 
\begin{align}
	\label{continuity_eulerian_final}
	\derivp{\overline{\rho}}{t}  
	+ \vec \nabla_\bot \cdot \left( \overline{\rho} \, \overline{\vec \vt}_\bot^\dag \right)
	&= 0 \, , \\
	\label{momentum_eulerian_final}
	\overline{\rho} \derivp{\overline{h}}{t}   
	+ \overline{\rho} \, \left( \overline{\vec \vt}_\bot^\dag \cdot \vec \nabla_\bot \right) \overline{h} 
	&+ \vec \nabla_\bot \cdot \left( \overline{\rho} \, \vec F \right) = \varpi \overline{X}_\phi \, , \\
	\label{energy_eulerian_final}
	\overline{\rho} \derivp{\overline{s}}{t} + 
	\overline{\rho}\left( \overline{\vec \vt}_\bot^\dag \cdot \vec \nabla_\bot \right) \overline{s} 
	&+ \vec \nabla_\bot \cdot \left( \overline{\rho}\, \vec G \right)
	= \overline{Q}  \, , 
\end{align}
where the components of the vectors $\vec F$ and $\vec G$ are given by
\begin{align}
	\label{waves_fluxes}
	F_r &= \varpi \, \overline{\vt_\phi^\prime \, \vt_r^\prime} + \frac{\overline{\psi} }{r} \, \derivp{\overline{h}}{\theta} \; , \quad 
	F_\theta = \varpi \, \overline{\vt_\phi^\prime \, \vt_\theta^\prime} - \overline{\psi} \, \derivp{\overline{h}}{r} \, , \\
	G_r &=  \overline{s^\prime \, \vt_r^\prime} + \frac{\overline{\psi} }{r} \, \derivp{\overline{s}}{\theta}  \; , \quad
	G_\theta =  \overline{s^\prime  \, \vt_\theta^\prime} - \overline{\psi} \, \derivp{\overline{s}}{r} \, .
\end{align}
Equation~\eqref{baroclinic_eulerian} is left unmodified. Note that $\vec F$ is often named to as the Eliassen-Palm flux because of the pioneer paper by Eliassen \& Palm (1961). Equations~\eqref{continuity_eulerian_final} to \eqref{energy_eulerian_final} seem very similar to Eqs.~\eqref{continuity_eulerian} - \eqref{energy_eulerian} but clarifies the exact role of the wave flux. To make clear the advantages of the TEM, it is useful to place ourselves in the limit of quasi-shellular approximation. 

\subsection{The TEM within the quasi-shellular approximation}

Shellular rotation, an often used approximation in stellar radiative zones,  assumes that an efficient horizontal transport of angular momentum is at work so that angular velocity is almost constant on isobars. Without discussing here the validity of this assumption (but see Maeder 2009 for a comprehensive discussion), it is a useful framework that permits us to exhibit the advantages of the TEM equations. 

Here we first remind the basic properties of this approximation. To do so, let us first introduce the following decomposition
\begin{align}
\label{shellular_omega}
\overline{\Omega} (r,\theta)= \Omega_0 \left(r \right) + \widehat{\Omega} \left(r,\theta\right) \, , 
\end{align}
with
\begin{align}
\Omega_0   = \frac{ \int_{0}^{\pi} \sin^3 \theta \; \overline{\Omega}(r,\theta)\; {\rm d} \theta}{ \int_{0}^{\pi} \sin^3 \theta \; {\rm d} \theta} = \frac{3}{4} \int_{0}^{\pi} \sin^3 \theta \; \overline{\Omega}(r,\theta)\; {\rm d} \theta ,
\end{align}
where, within the shellular approximation, $\Omega_0 \gg \widehat{\Omega}$. 
The scalar quantities are developed as 
\begin{align}
\label{def_scalar} 
\overline{X}(r,\theta) &=  \left< X \right> \left(r \right) + \widehat{X} \left(r,\theta\right) \, , 
\end{align} 
with
\begin{align}
\left< X \right>   = \frac{ \int_{0}^{\pi} \sin \theta \; \overline{X}(r,\theta) \, {\rm d} \theta}{ \int_{0}^{\pi} \sin \theta \, {\rm d} \theta} = \frac{1}{2} \int_{0}^{\pi} \sin \theta \; \overline{X}(r,\theta) \, {\rm d} \theta \, .
\end{align}
For slow to moderate rotation, we assume that $\widehat{X} \ll  \left< X \right>$. 

Following these assumptions,  and to have more insight about the advantage of using the TEM formulation, it is useful to emphasize that in stars, we are generally in the situation where the isentropic and isopycnal surfaces are nearly parallel. Therefore, we are in the situation for which we have 
\begin{align}
\label{approx_shellular}
\left\vert \frac{1}{r} \derivp{\overline{s}}{\theta} \right\vert \ll \left\vert \derivp{\overline{s}}{r} \right\vert \, , 
\end{align}
so that the stream function reduces to 
\begin{align}
\label{def_psi_general_simplified}
\overline{\psi} \simeq  \left(\derivp{\overline{s}}{r}\right)^{-1} \, \overline{s^\prime \vt_\theta^\prime}   \, , 
\end{align}
which is the same as what is found within the quasi-geostrophic approximation. Then, we immediately show that $G_\theta = 0$ and that 
\begin{align}
\label{residual_flux}
G_r \simeq  \overline{s^\prime \, \vt_r^\prime} \, . 
\end{align}

To clarify the advantage of the TEM, let us consider the perturbed equation of the entropy by using \eq{energy} and \eq{energy_eulerian}. It gives  
\begin{align}
\derivp{s^\prime}{t}  + \left(\overline{\vec \vt}\cdot \vec \nabla\right) s^\prime +  \left(\vec \vt^\prime\cdot \vec \nabla\right) \overline{s} = \tilde{Q}^\prime + O\left(a_w^2\right) \, , 
\end{align}
where $\tilde{Q} \equiv Q/\rho$. Multiplying by $s^\prime$ and performing azimuthal average then gives
\begin{align}
\frac{1}{2}\derivp{\overline{s^{\prime 2}}}{t}  + \frac{1}{2} \overline{\left(\overline{\vec \vt}_\bot \cdot \vec \nabla_\bot \right) s^{\prime 2}} +  \overline{s^\prime \vec \vt^\prime} \cdot \vec \nabla_\bot \overline{s} = \tilde{Q}^\prime + O\left(a_w^3\right) \, . 
\end{align}
The first term corresponds to the growth rate of the wave, which can be assumed to be small except in the driving regions and the second term is of the order of $O\left(a_w^4\right)$ because meridional circulation is $O\left(a_w^2\right)$ as it is predominantly wave-driven. Finally, assuming $\tilde{Q}^\prime=0$ one gets 
\begin{align}
  \overline{s^\prime \vec \vt^\prime} \cdot \vec \nabla_\bot \overline{s} = 0 + O\left(a_w^3\right) \, , 
\end{align}
which means that the entropy wave flux is nearly perpendicular to the mean entropy gradient. In other words, the flux is almost horizontal in the quasi-shellular approximation because using \eq{approx_shellular} one gets
\begin{align}
\frac{ \overline{s^\prime \vt_\theta^\prime}}{ \overline{s^\prime \vt_r^\prime}}  \ll 1 \, . 
\end{align}

Coming back to the TEM formalism and Eqs.~(\ref{continuity_eulerian_final}) to (\ref{energy_eulerian_final}), means that most of the wave heat flux has been removed from the entropy equation and the remaining flux (Eq.~\ref{residual_flux}) is a residual. Consequently, we end up with 
\begin{align}
\label{momentum_eulerian_final_approx}
\overline{\rho} \derivp{\overline{h}}{t}   
+ \overline{\rho} \, \left( \overline{\vec \vt}_\bot^\dag \cdot \vec \nabla_\bot \right) \overline{h} 
&+ \vec \nabla_\bot \cdot \left( \overline{\rho} \, \vec F \right) = \varpi \overline{X}_\phi  \\
\label{energy_eulerian_final_approx}
\overline{\rho} \derivp{\overline{s}}{t} + 
\overline{\rho}\left( \overline{\vec \vt}_\bot^\dag \cdot \vec \nabla_\bot \right) \overline{s} 
&
= \overline{Q}  \, , 
\end{align}
Therefore, the TEM formalism allows us to gather in a single equation and in the single term $\vec F$ both the wave momentum and wave heat fluxes. It explicitly shows that wave momentum fluxes and wave heat fluxes do not influence the mean flow separately, but only in the combination given by $\vec F$.

Finally, when $\vec F$ is specified, the residual velocity $\overline{\vec \vt}_\bot^\dag$ becomes part of the solution of Eqs.~\eqref{momentum_eulerian_final_approx} and \eqref{energy_eulerian_final_approx} together with Eq.~\eqref{baroclinic_eulerian}. They are  strictly equivalent to Eqs.~\eqref{continuity_eulerian} - \eqref{energy_eulerian}, but  the TEM equations enable distinguishing the advective and diffusive parts of the wave heat flux and incorporating the advective component in the mean velocity field. More importantly, Andrews \& McIntyre (1978a) showed that $\vec \nabla \cdot  \left(\overline{\rho} \vec F\right)$ only depends on wave dissipation and non-steady terms. It is the \emph{non-acceleration theorem}. Therefore, the TEM makes the adiabatic and non-adiabatic contributions of waves more explicit, the latter being only able to modify the mean flow. 

\subsection{The TEM as seen by the Generalized Lagragian Mean (GLM) formalism}
\label{sect5}

In this section, we very briefly discuss the Generalized Lagrangian Mean (GLM) formalism, which has been developed by Andrews \& McIntyre (1978b,c). More details on this theory can be found in Craik (1988),  Grimshaw (1984), and Buhler (2009). Also in stellar physics, this formalism was first used by Lee (2013) and in later papers such as in Lee et al. (2016) for massive stars. 

\begin{figure}[!]
	\centering
	\includegraphics[width=12cm]{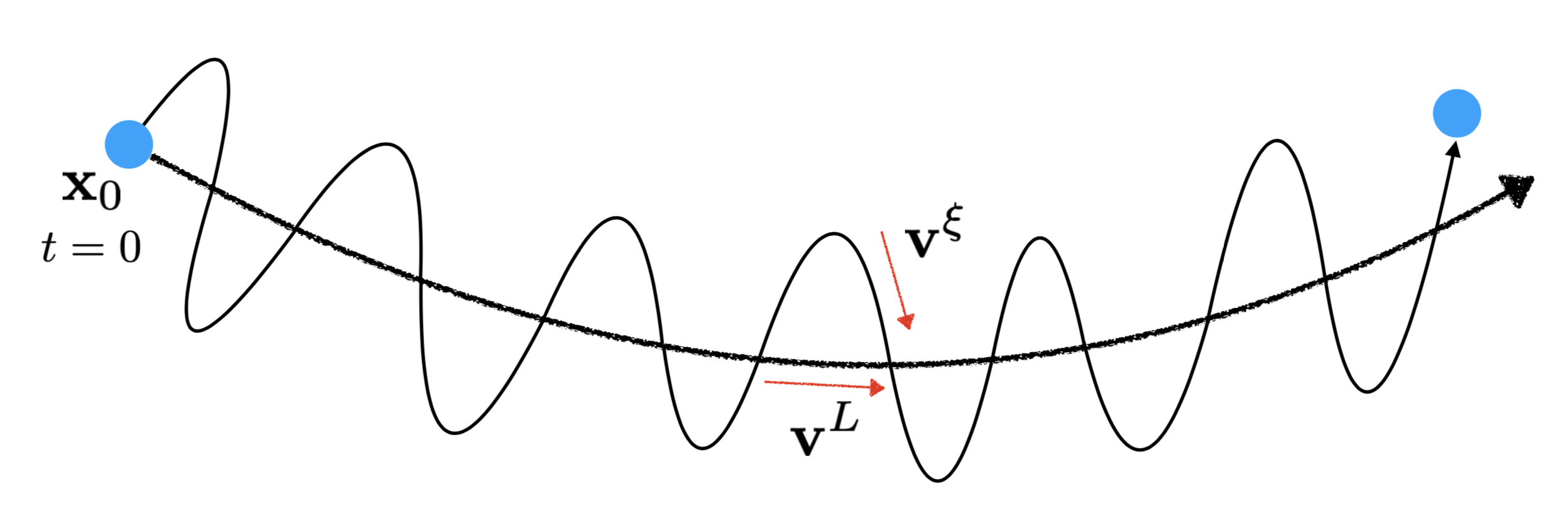}
	\caption{Schematic sketch of a fluid particle trajectory with an initial position $\vec X_0 (t=t_0)$. $\overline{\vt}^{L}$ and $\vec \vt^{\xi}$ are the Lagrangian mean as defined by \eq{glm_def_v_lag_mean} and the Lagagian velocity as defined by \eq{glm_def_v_lag}. \label{fig_4}}
\end{figure}

The GLM formalism can be understood as a generalization of the TEM formalism for finite amplitude waves but, even for small wave amplitudes, it provides an enlightening theoretical framework for understanding the wave/mean-flow interactions and the TEM. Let us first define the Lagrangian velocity following the wave displacement
\begin{equation}
\label{glm_def_v_lag}
\vec \vt^{\xi} \equiv \vec \vt \left(\vec x + \vec\xi \left(\vec x,t\right), t \right) \, , 
\end{equation}
where $\vec \xi$ is the wave displacement. The main idea underlying the GLM is to adopt a mixed Eulerian-Lagragian approach in which mean Lagrangian velocity $\overline{\vec \vt^{\xi}}$ is defined by
\begin{equation}
\label{glm_def_v_lag_mean}
\overline{\vt}^{L} \equiv \overline{\vec v\left(\vec x + \vec\xi \left(\vec x,t\right), t \right)} \, .
\end{equation}
Here, the Eulerian average can be either a spacial, temporal, or ensemble average and as such the GLM is a very general theory. A schematic sketch of the two above-defined velocities is shown in Fig.~\ref{fig_4} for a fluid particle trajectory. 

To illustrate one of the advantage of the GLM, let us consider the conservation of specific entropy. It can be written such as 
\begin{equation}
\label{entropy_main}
\frac{ D s}{D t} =Q \, .
\end{equation}
Equation (\ref{entropy_main}) can then be averaged using the classical Eulerian average so that it gives 
\begin{align}
\frac{ D \overline{s}}{D t} - \overline{Q} = -\overline{\vec v^\prime \cdot \vec \nabla s^\prime} \, , 
\end{align}
which immediately shows that the mean entropy is not conserved (even if $\overline{Q} = 0$) due to a second-order wave related term. In sharp contrast, by adopting the GLM, one can easily show that 
\begin{align}
\label{entropy_GLM}
\overline{D}^L \overline{s}^L  = \overline{Q}^L \, , 
\end{align}
where the Lagrangian derivative is defined by 
\begin{equation}
\label{glm_res}
\overline{D}^L \equiv \derivp{}{t} + \vec v^{L} \cdot \vec \nabla \, .
\end{equation}
Equation (\ref{entropy_GLM}) demonstrates that using the GLM permits the mean equation to conserve its conservative form. Then, for $\overline{Q}^L =0$, the mean entropy (here $\overline{s}^L$) is conserved. Note that, as for the TEM but here exactly, the only wave contributions to the mean equation appear in the momentum equation. In the limiting case $X=0$ and $Q=0$ this only contribution appears through the pseudo-momentum. For adiabatic waves, this is a conserved quantity as shown by Andrews \& McIntyre (1978b,c) thus demonstrating the non-acceleration theorem. 

When compared to the TEM, this is an exact result and there is no residual term compared to \eq{energy_eulerian_final}. To understand this, it must be realized that the difference between the Lagrangian and Eulerian quantities are the Stokes corrections. For a scalar quantity, the Stokes correction is defined by
\begin{align}
\overline{\phi}^S \equiv \overline{\phi}^L - \overline{\phi} \, .
\end{align}
Therefore, the wave-related term in \eq{glm_res} can be written
\begin{align}
\overline{\vec v^\prime \cdot \vec \nabla s^\prime} = \overline{\vec \vt}^S \cdot \vec \nabla \overline{s} + \left(\derivp{}{t} + \overline{\vec \vt}^S \cdot \vec \nabla  \right) \overline{s}^S - \overline{Q}^S \, . 
\end{align}
The latter equation permits us to understand the difference between \eq{entropy_main} and \eq{entropy_GLM} as Stokes corrections are  second-order quantities in term of wave amplitude. Furthermore, it also gives us an interpretation for the advective part of the divergence of the wave heat flux appearing in \eq{split_div_flux} within the TEM. This velocity can be understood as due to the Stokes correction. We also stress that in the case $Q=0$ and if the wave is adiabatic one has
\begin{align}
\overline{\vec \vt}_\bot^\dag = \overline{ \vt}^L \, , 
\end{align}
which means that GLM and residual circulations are the same in the limiting case of $Q=0$ and adiabatic waves. 
 
\section{Application to mixed modes in the low-mass evolved stars}
\label{sect4}

In this section, based on the work of Belkacem et al. (2015a,b), we illustrate how the TEM can be used to estimate the transport of angular momentum by waves. This work focused on mixed modes but can also be easily generalized for progressive waves. 

\subsection{The problem of angular mometum reditribution in low-mass evolved stars}

\begin{figure}[!]
	\centering
	\includegraphics[width=14cm]{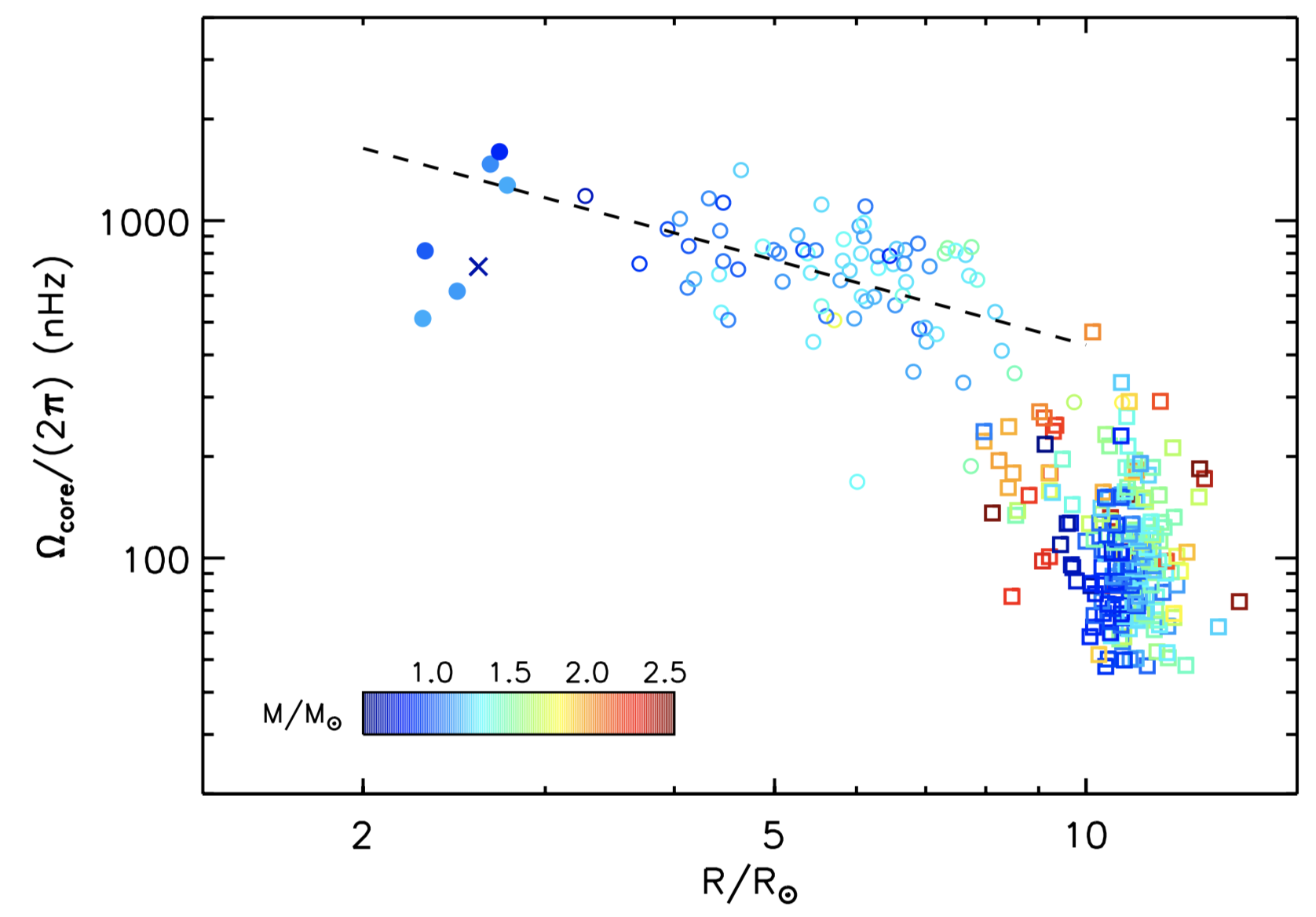}
	\caption{Rotation rate of the core as a function of the stellar radius. The open dots and open squares correspond to the RGB and clump stars, respectively, investigated by Mosser et  al. (2012). The filled symbols indicate the subgiant stars studied by Deheuvels et al. (2012,2014). Figure from Deheuvels et al. (2015) \label{fig_1}}
\end{figure}

The \emph{Kepler}  (Borucki et al. 2010) space-borne mission permitted us to unveil the rotation of the innermost layers of evolved low-mass stars. Based on seismic measurements, Beck et al. (2012), Deheuvels et al. (2012, 2014)  brought  constraints on the rotation profiles of a handful of subgiant stars and concluded that the core of subgiant stars spins up, while their envelope decelerates. Due to the contraction of their core and the expansion of their envelope, while such a result could appear as not being surprising, the problem is that local conservation of angular momentum would have produced a much more pronounced differential rotation. Therefore, a physical mechanism (which is to be identified) is needed for extracting angular momentum and thus explaining the observations.  For more evolved stars, red giants, Mosser et al. (2012) and Gehan et al. (2018) analysed a sample of hundreds of red-giant stars observed by \emph{Kepler} and found that, surprisingly, the mean core rotation rate decreases significantly during the red-giant phase. The actual observational picture is summarized by Fig.~\ref{fig_1}. All those observational constraints emphasize the need of angular momentum redistribution between the core and the envelope. However, current models of red-giant stars including angular momentum redistribution processes are unable to explain such low core rotation rates in subgiant and red giant stars. They are also unable to explain the deceleration of the core during the ascent of the red-giant branch. 

Consequently, the quest for a physical mechanism able to extract angular momentum from the core of evolved stars is currently under way. Meridional circulation and shear instabilities have been shown to be inefficient (e.g.,  Eggenberger et al. 2012, Marques et al. 2013). The same conclusion was reached by Cantiello et al. (2014) for a magnetic field generated through the Tayler-Spruit dynamo (Spruit 1999, 2002). However, Fuller et al. (2019) investigate the magnetic Tayler instability and argue they were able to  reproduce the nearly rigid rotation of main sequence stars and the core rotation rates of low-mass red giants. For internal gravity waves, it has been found that they are unable to explain  the slow-down of red-giants (see Fuller et al. 2014 and Pin\c{c}on et al. 2017). In contrast, Pin\c{c}on et al. (2017) had demonstrated that it is able to explain the differential rotation as observed in subgiants. 

In this context, Belkacem et al. (2015a,b) investigated the ability of mixed modes to extract angular momentum within the TEM formalism. The aim was to estimate the influence of mixed modes on the angular momentum evolution compared to the structural effect of contraction of the core. To do so, they  
started from \eq{momentum_eulerian_final_approx} and after meridional average, the equation governing angular momentum evolution can be written as 
\begin{align}
\label{shellular_omega_r}
&\left< \rho \right> \deriv{\left( r^2 \Omega_0 \right)}{t}   
= -\frac{1}{r^2} \derivp{}{r} \left( r^2  \mathcal{F}_{\rm waves}  \right) \equiv \dot{J} \,, 
\end{align}
where the symbol $\left< \right>$ denotes the horizontal average (\emph{i.e.} azimuthal and meridional). The wave flux is defined by
\begin{align}
\label{shellular_omega_r_wave}
\mathcal{F}_{\rm waves} =  \left< \rho \right>\left< \varpi \left[ \overline{\vt_\phi^\prime \, \vt_r^\prime} + 2 \cos \theta \,  \Omega_0 \, \overline{\vt_\theta^\prime s^\prime} \left(\deriv{\left< s\right>}{r}\right)^{-1} \right] \right> \, ,
\end{align}
where $\varpi = r \sin \theta$, the prime denotes perturbations associated with the non-radial oscillations, so that $\vt_\phi^\prime$, $\vt_r^\prime$, $\vt_\theta^\prime$ are the azimuthal, radial, and meridional component of the wave velocity field, and $s^\prime$ the wave Eulerian perturbation of entropy.  We have also introduced the Lagrangian derivative ${\rm d} / {\rm d}t =  \partial / \partial t +  \dot{r} \, \partial / \partial r$. 

\subsection{Computing the wave field}
\label{wave_approx}

To go further and to quantify the transport of angular momentum by mixed modes in the radiative region of evolved low-mass stars (i.e. subgiants and red giants), one has first to compute the wave field and thus to estimate the wave flux as given by \eq{shellular_omega_r_wave}. As these modes have a dual nature; they behave as acoustic modes in the upper layers and as gravity modes in the inner layers, several assumptions can then be adopted by using shellular rotation and focusing on the inner radiative regions. These restrictions allow the use of several approximations to describe the wave field: 
\begin{enumerate}
	\item \emph{The quasi-adiabatic approach:} It consists in neglecting the difference between adiabatic and non-adiabatic eigenfunctions in the full wave equations. 
	This approximation is valid when the local thermal time-scale is much longer than the modal period. This is the case in the radiative region of evolved low-mass  stars. 
	\item \emph{The low-rotation limit:} the modal period is assumed to be much shorter than the rotation period. This is justified by inferences of the rotation in the core of subgiants (Deheuvels et al. 2012, 2014) and red giants Mosser et al. (2012) using seismic constraints from \emph{Kepler}. 
	\item \emph{The asymptotic limit:} an asymptotic description for gravity modes (e.g., Dziembowski et al. 2001; Godart et al. 2009) is valid for mixed modes in the inner radiative region of subgiants and red giants (e.g., Goupil et al. 2013). 
\end{enumerate}

The next requirement is the determination of mode amplitudes. This is a fundamental point since it determines the amount of angular momentum transported by mixed modes. To do so, there are mainly two approaches. The first  is based on a full non-adiabatic computation including a time-dependent treatment of convection. This procedure is time-consuming, however, and still suffers from theoretical uncertainties (see Dupret et al. 2009, Grosjean et al. 2014, for details). 
The second is based on recent CoRoT and \emph{Kepler} observations that  made it possible to establish scaling relations that provide mode amplitudes versus global stellar parameters (e.g., Mosser et al. 2012, Samadi et al. 2012). Belkacem et al. (2015b) followed the latter approach as it provides more reliable results. 

\subsection{Efficiency of angular momentum extraction}
\label{mixed_efficiency}
Using the simplifying assumptions as described in Sect.~\ref{wave_approx} and after quite involving calculations, Belkacem et al. (2015a,b) computed the divergence of the wave flux appearing in \eq{shellular_omega_r_wave}. It led to the conclusion that prograde modes ($m <0$) extract angular momentum in the regions near the maximum of the buoyancy frequency and the maximum of the rotation rate, therefore slowing down the core. Conversely, the retrograde modes ($m>0$) tend to spin up the core. 

However, the net angular momentum flux is related to the asymmetry between prograde and retrograde modes. Therefore, it is worthwhile computing the net contribution of prograde and retrograde modes for a given angular degree. A simplified, but still accurate, expression is given by (see Belkacem et al. 2015a for details) 
\begin{align}
\label{exp_simplified_j}
\dot{J}(\ell,-\vert m \vert) + \dot{J}(\ell,\vert m \vert) \approx 2 \vert m \vert^2 \rho k_r^2 a_{\ell,\vert m \vert}^2  \vert \xi_r^{\ell,\vert m \vert } \vert^2 \left(\frac{\Omega_0}{\sigma_R}\right) \left(\frac{N^2}{\sigma_R}\right) \, \alpha \, , 
\end{align}
where $m$ is the azimuthal order, $k_r$ the radial wave number, $a_{\ell,\vert m\vert}$ the mode amplitude, $\xi_r^{\ell,\vert m\vert}$ the radial eigen-function, $\sigma_R$ the pulsational frequency, and 
\begin{align}
\label{alpha}
\alpha = -\frac{L}{4 \pi r^2 \rho T} \left( \frac{\nabla_{\rm ad}}{\nabla}-1 \right) \, \left(\deriv{s}{r}\right)^{-1} 
\end{align}
with $L$ the luminosity, $T$ the temperature, $\nabla$ and $\nabla_{\rm ad}$ the temperature gradient and its adiabatic counterpart, and $s$ the entropy. 

Because the term $\alpha$ (see Eq.~\ref{alpha}) is negative, \eq{exp_simplified_j} is also negative so that the sum of the contributions of prograde and retrograde mixed modes implies an extraction of angular momentum. Therefore, the collective effect of mixed modes is to decrease the mean angular momentum and thus to slow down the core rotation of the star. 

\begin{figure}[h]
	\centering
	\includegraphics[width=8cm]{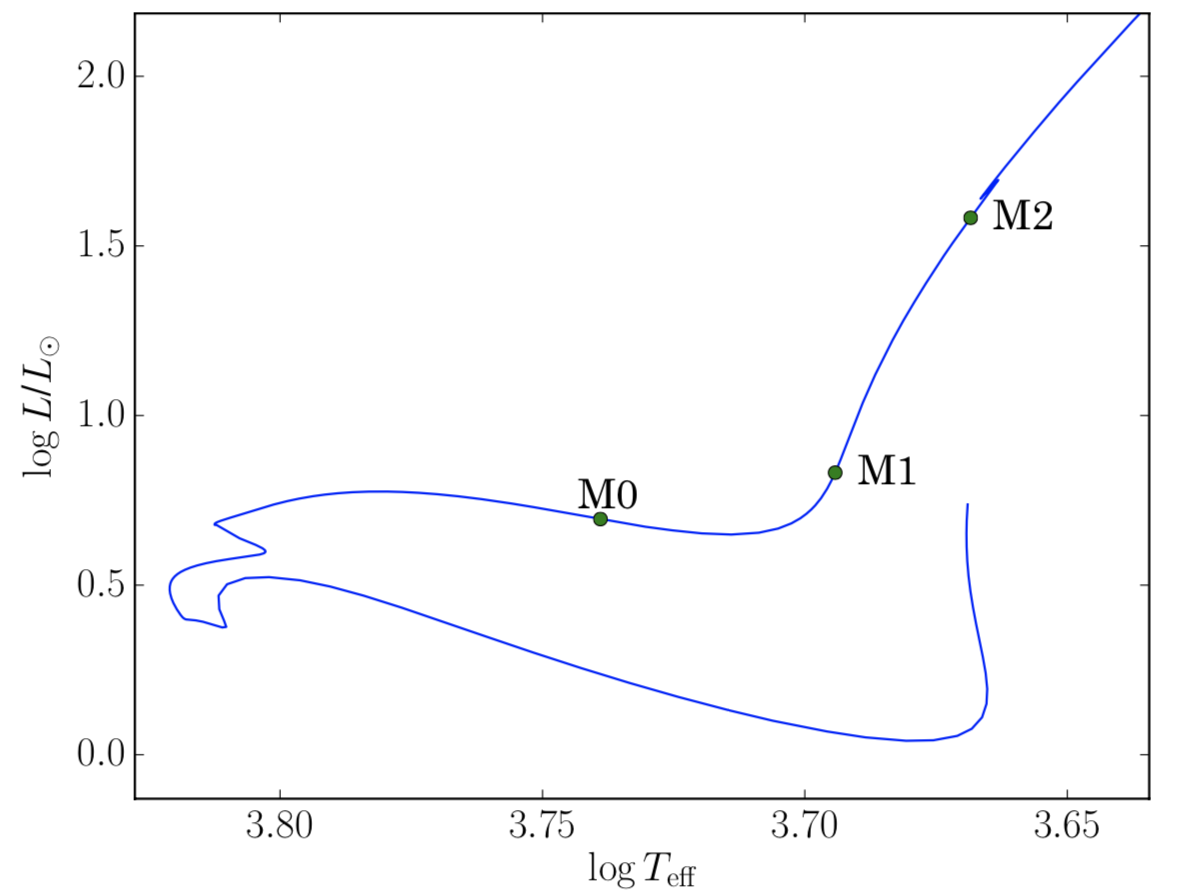}
	\includegraphics[width=8.5cm]{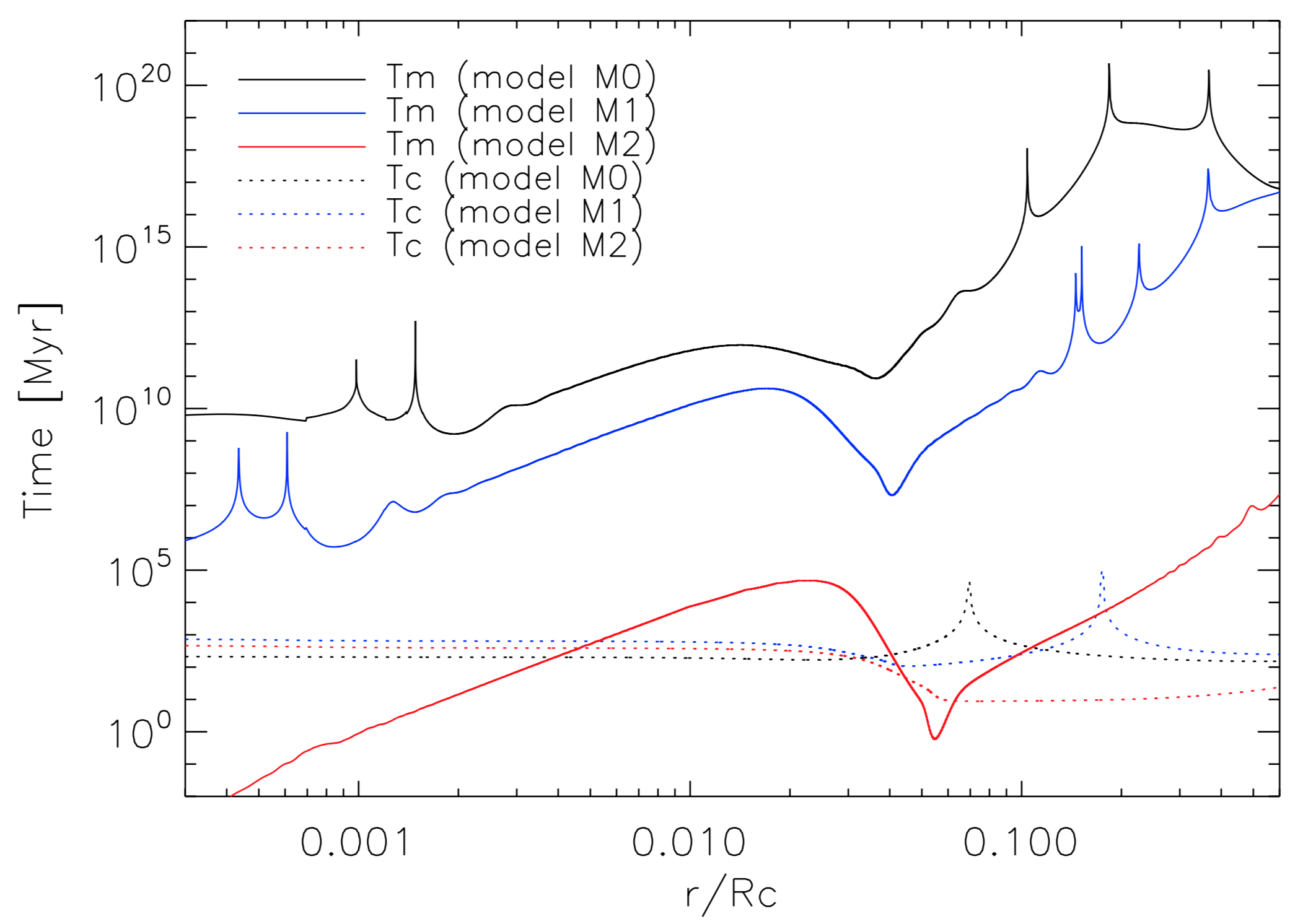}
	\caption{\emph{Right panel:} Evolutionary track on the HR diagram of a $1.3 \, M_\odot$ model showing the location of the three considered models in Sect.~\ref{mixed_efficiency}. Figure from Belkacem et al. (2015b). \emph{Left panel:}  Timescales versus normalised radius (i.e., normalised by the radius of the base of the convective envelope) for models M0, M1, and M2. The solid lines correspond to the timescale associated with the transport of angular momentum by mixed modes (see Eq.~\ref{time_modes})) and the dotted lines correspond to the timescale associated with the contraction of the star (see Eq.~\ref{time_spin}).  Figure from Belkacem et al.(2015b) \label{fig_2}}
\end{figure}

To go further and to assess the efficiency of the transport of angular momentum, four benchmark models have been chosen as depicted in Fig.~\ref{fig_2} (right panel). For those models, it is useful to define two characteristic time-scales, namely; 
\begin{itemize}
	\item the timescale associated to the efficiency of the transport of angular momentum by mixed modes, defined as
	\begin{align}
	\label{time_modes}
	T_m^{-1} = \left\vert \frac{\dot{J}}{\rho r^2 \Omega_0} \right\vert \, .
	\end{align}
	\item the timescale associated to the contraction of the star, defined as
	\begin{align}
	\label{time_spin}
	T_c^{-1} =\left\vert -\frac{1}{\rho r^4 \Omega_0} \derivp{}{r} \left(\rho r^4 \Omega_0 \, \dot{r}\right) \right\vert  \approx \left\vert - \frac{\dot{r}}{r} \right\vert  \, ,
	\end{align}
	where $\dot{r}={\rm d} r / {\rm d} t$.  
\end{itemize} 
Both timescales are shown in Fig.~\ref{fig_2} (left panel). In all the radiative regions of models M0 and M1, the extraction of angular momentum remains negligible compared to the star contraction. In contrast, for the model M2, the timescale of angular momentum extraction is of the same order of magnitude as  the contraction timescale and even lower in the hydrogen shell burning region for which the contraction is maximal. Therefore, one can conclude that the extraction of angular momentum by mixed modes is negligible in subgiants and early red giants while it becomes important  in the hydrogen burning shell in stars higher on the red-giant branch. In such cases, mixed modes are able to counterbalance the spin-up due to the star contraction and can thus enforce a spin-down in those layers. The amount of angular momentum extracted by mixed modes increases with the evolutionary stage of the star. This effect is the result of several factors. As shown in Fig.~\ref{fig_2}, mode amplitudes increase from models M0 to M2 and thus more energy is available to transport angular momentum. Moreover, the buoyancy frequency also increases and so does the radial wave number in the gravity-mode cavity. As the energy exchanges between modes and the background is proportional to the radial wave number, the amount of angular momentum extracted also increases. Finally, the number of mixed modes between two p-dominated modes significantly increases between M0 and M2. 

\section{Concluding remarks}
\label{sect6}

In this article, we tried to show that investigating the transport of angular momentum by waves in stars demands to address the more wide issue of the interaction between the waves and the mean-flow. Indeed, the classical zonal or azimuthal average introduces wave-related terms in both the mean momentum and mean energy equations. Both are to be considered as they do not act independently because they are connected through meridional circulation. We therefore described the Transformed Eulerian formalism which ensures a proper modelling of the wave/mean-flow interactions. This formalism, adapted for small wave-amplitudes, allows to properly address the problem. This has been illustrated in the case of angular momentum transport by mixed modes in low-mass red giant stars. A more general theory, the Generalized Lagrangian Mean formalism has also been briefly introduced as it provides a nice framework for understanding the problem. 

Finally, we stress that considering only the mean momentum equation using a classical Eulerian average is a too naive approach which does not permit to grasp the physics of the problem. It can potentially leads to incorrect results because the coupling with the wave-related terms in the energy equation as ensured by meridional circulation is overlooked. Therefore, it would  be desirable in the future to reanalyze previous works that only considered the mean momentum equation in the TEM and GLM frameworks.

% USE A SECTION WITHOUT NUMBER FOR THE ACKNOWLEDGEMENTS
%
\section*{Acknowledgements}
I thank Marc-Antoine Dupret for his invitation to give a talk to the Liege conference, dedicated to Arlette Noels 75th birthday, and for his careful reading of the manuscript. I was honoured to spend several years in her research group and it was a very nice period when I have learnt a lot thanks to Arlette. I thus sincerely thanks Arlette Noels for our enlightening discussions and her constant support. I finally acknowledge financial support from the “Programme National de Physique Stellaire” (PNPS) of CNRS/INSU.

%
% BEGIN THE REFERENCE LIST WITH \beginrefer
% USE \refer BEFORE THE REFERENCES AND BEGIN A NEW PARAGRAPH AFTER THE 
% REFERENCE !
% DO NOT FORGET TO END THE LIST WITH \endrefer
% 
%
% INSTRUCTIONS FOR BIBLIOGRAPHY:
% ==============================
% - DON'T USE THE & SYMBOL
% - USE INITIALS FOR FIRST AND MIDDLE NAMES, AND SPECIFY FULL FAMILY NAME (see examples below)
% - NO COMMA BETWEEN NAME AND INITIALS
% - USE COMMA BETWEEN DIFFERENT AUTHORS NAMES
% - NO COMMA AFTER THE LAST AUTHOR NAME
% - FOR LONG AUTHOR LISTS, SPECIFY THE FIRST 3 AUTHORS FOLLOWED BY 'et al.', WITH NO COMMA BEFORE AND AFTER 'et al.'
% - INSERT A BLANK SPACE BETWEEN MULTIPLE INITIALS
% - USE STANDARD JOURNAL ACRONYMS FREQUENTLY USED IN MAIN ASTROPHYSICS JOURNAL
% - SORT REFERENCES BY ALPHABETICAL ORDER OF FIRST AUTHOR NAMES
% - MULTIPLE REFERENCES WITH THE SAME FIRST AUTHOR SHOULD BE SORTED BY CHRONOLOGICAL ORDER

\footnotesize

\beginrefer

\refer Ando H. 1983, PASJ, 35, 343

\refer Ando H. 1986, A\&A, 163, 97

\refer Andrews D. G. and McIntyre M. E. 1976, Journal of Atmospheric Sciences, 33, 2031 

\refer Andrews D. G. and McIntyre M. E. 1978a, Journal of Atmospheric Sciences, 89, 609

\refer Andrews D. G. and McIntyre M. E. 1978b, Journal of Fluid Mechanics, 35, 175

\refer Andrews D. G. and McIntyre M. E. 1978c, JJournal of Fluid Mechanics,  89, 647

\refer Andrews D. G., Holton J. R., and Leovy C. B. 1987, Middle atmosphere dynamics (Academic Press) 

\refer Beck P. G., Montalban J., Kallinger T., et al. 2012, Nature, 481, 55

\refer Borucki W. J., Koch D., Basri G., et al. 2010, Science, 327, 977

\refer Bühler O., 2009, \emph{Waves and Mean Flows}, Cambridge University Press, Cambridge, UK, ISBN 978-0-521-86636-1.

\refer Cantiello M., Mankovich C., Bildsten L., Christensen-Dalsgaard J., and Paxton B. 2014, ApJ, 788, 93

\refer Craik A. D. D. 1988, \emph{Wave Interactions and Fluid Flows}, Cambridge University Press, ISBN 0521368294. 

\refer Charbonnel C., Talon S., 2005, Science, 309,  5744

\refer Deheuvels S., Dogan G., Goupil M. J., et al. 2014, A\&A, 564, A27

\refer Deheuvels S., Garc ia, R. A., Chaplin, W. J., et al. 2012, ApJ, 756, 19

\refer Dintrans B. and Rieutord M. 2001, MNRAS, 324, 635

\refer Dupret M.-A., Belkacem K., Samadi R. et al. 2009, A\&A, 506, 57

\refer Dziembowski W. A., Gough D. O., Houdek G., and Sienkiewicz R. 2001, MNRAS, 328, 601

\refer Eggenberger P., Montalban J.,  and Miglio A. 2012, A\&A, 544, L4

\refer Eliassen  A.  N.,  and  Palm E. ,  1961, Geof. Pub., 22, 1-23

\refer Fuller J., Lecoanet D., Cantiello M., and Brown B. 2014, ApJ, 796, 17

\refer Fuller J., Piro A. L., Jermyn A. S. 2019, MNRAS, 485,  3661 

\refer Gehan C., Mosser B., Michel, E., Samadi, R., Kallinger T. 2018, A\&A,  616, A24

\refer Godart M., Noels A., Dupret M.-A., and Lebreton Y. 2009, MNRAS, 396, 1833

\refer Goupil M. J., Mosser B., Marques J. P. et al. 2013, A\&A, 549, A75

\refer Grimshaw R. 1984, Ann. Rev. Fluid. Mech., 16, 11

\refer Grosjean M., Dupret M.-A., Belkacem K. et al. 2014, A\&A, 572, A11

\refer Holton, J. R., 1992, \emph{An introduction to dynamic meteorology}, International geophysics series, San Diego, New York: Academic Press, 3rd ed.

\refer Ishimatsu H., and Shibahashi H. 2013, ASP Conference Proceedings, 479,  325

\refer Kumar P., Talon S., and Zahn J.-P. 1999, ApJ, 520, 859

\refer Lee U. and Saio H. 1993, MNRAS, 261, 415

\refer Lee U. 2007, ASP Conf. Ser., 361, 45

\refer Lee U. 2013, PASJ, 65, 122

\refer Lee U., Neiner C., and Mathis S. 2014, MNRAS, 443, 1515

\refer Lee U., Mathis S.,  and Neiner C. 2016, MNRAS, 457, 2445

\refer Maeder A., 2009, \emph{Physics, Formation and Evolution of Rotating Stars}, Astronomy and Astrophysics Library. Springer Berlin Heidelberg.  ISBN 978-3-540-76948-4.

\refer Marques J. P., Goupil M. J., Lebreton Y., et al. 2013, A\&A, 549, A74

\refer Mathis S. 2013, Lecture Notes in Physics, 865, 23

\refer Mosser B., Goupil M. J., Belkacem K., et al. 2012, A\&A, 548, A10

\refer Pantillon F. P., Talon S., and Charbonnel, C. 2007, A\&A, 474, 155

\refer Pinçon C., Belkacem K., Goupil M. J., Marques J. P. 2017, A\&A, 605, 31

\refer Press W. H. 1981, ApJ, 245, 286

\refer Samadi R., Belkacem K., Dupret M.-A. et al. 2012, A\&A, 543, A120

\refer Schatzman E., 1993, A\&A, 279, 431

\refer Shibahashi H., 2014, in IAU Symposium, 301, 1

\refer Spruit H. C., 2002, A\&A, 381, 923

\refer Spruit H.C., 1999, A\&A, 349, 189 

\refer Talon S., Charbonnel C., 2003, A\&A, 405, 1025

\refer Talon S., Charbonnel C., 2005, A\&A, 440, 3 

\refer Townsend R. 2014, in IAU Symposium, 301, 153–160

\refer Townsend R. H. D., Goldstein J., and Zweibel E. G 2018, MNRAS, 475, 879

\refer Townsend R., and MacDonald J. 2008, in IAU Symposium, 250, 161

\refer Unno, W., Osaki, Y., Ando, H., Saio, H., \& Shibahashi, H. 1989, Nonradial oscillations of stars (University of Tokyo Press, 1989, 2nd ed.)

\refer Zahn J.-P., Talon S., and Matias, J. 1997, A\&A, 322, 320

\endrefer           

\end{document}